\documentclass[reprint,amsmath,amssymb,aps,superscriptaddress]{revtex4-1}

\usepackage{graphicx}
\usepackage{dcolumn}
\usepackage{bm}

\begin{document}

\title{Selecting for adaptability in materials design}
\title{Design principles for adaptability revealed by time-varying optimization}
\title{Design principles for adaptability revealed by transient optimization}
\title{Design principles for adaptable materials revealed by incomplete optimization}
\title{Design principles for adaptable materials revealed by incomplete optimization}
\title{Design principles for adaptability revealed by incomplete optimization}
\title{Adaptable materials through incomplete design}
\title{Adaptable materials revealed by repeated incomplete design}
\title{Adaptable materials revealed by repeated incomplete design}
\title{Non-equilibrium training protocols for adaptable materials}
\title{Learning to learn: Non-equilibrium design protocols for adaptable materials}


\author{Martin J. Falk*}
\affiliation{Department of Physics, The University of Chicago, Chicago, IL 60637}

\author{Jiayi Wu*}
\affiliation{Department of Physics, The University of Chicago, Chicago, IL 60637}

\author{Ayanna Matthews}
\affiliation{Graduate Program in Biophysical Sciences, The University of Chicago, Chicago, IL 60637}

\author{Vedant Sachdeva}
\affiliation{Graduate Program in Biophysical Sciences, The University of Chicago, Chicago, IL 60637}

\author{Nidhi Pashine}
\affiliation{School of Engineering and Applied Science, Yale University, New Haven, CT 06511}

\author{Margaret Gardel}
\affiliation{Department of Physics, The University of Chicago, Chicago, IL 60637}
\affiliation{James Franck Institute, The University of Chicago, Chicago, IL 60637}
\affiliation{Institute for Biophysical Dynamics, The University of Chicago, Chicago, IL 60637}
\affiliation{Pritzker School of Molecular Engineering, The University of Chicago, Chicago, IL 60637}

\author{Sidney Nagel}
\affiliation{Department of Physics, The University of Chicago, Chicago, IL 60637}
\affiliation{James Franck Institute, The University of Chicago, Chicago, IL 60637}

\author{Arvind Murugan}
\affiliation{Department of Physics, The University of Chicago, Chicago, IL 60637}
\affiliation{James Franck Institute, The University of Chicago, Chicago, IL 60637}

\begin{abstract}
Evolution in time-varying environments naturally leads to adaptable biological systems that can easily switch functionalities. Advances in the synthesis of environmentally-responsive materials therefore open up the possibility of creating a wide range of synthetic materials which can also be trained for adaptability. We consider high-dimensional inverse problems for materials where any particular functionality can be realized by numerous equivalent choices of design parameters. By periodically switching targets in a given design algorithm, we can teach a material to perform incompatible functionalities with minimal changes in design parameters. We exhibit this learning strategy for adaptability in two simulated settings: elastic networks that are designed to switch deformation modes with minimal bond changes; and heteropolymers whose folding pathway selections are controlled by a minimal set of monomer affinities. The resulting designs can reveal physical principles, such as nucleation-controlled folding, that enable adaptability.
\end{abstract}

\maketitle



Considered as materials, biological systems are striking in their ability to perform many individually demanding tasks in contexts that can often change over time.
This success can be attributed to ``meta-properties'' like modularity\cite{kashtan2005spontaneous,Parter2008-ez,hemery2015evolution,kashtan2007varying,lipson2002origin}, robustness\cite{wagner2008robustness}, plasticity for learning\cite{gupta2021embodied}, and multifunctionality\cite{du2020inverse,rocks2019limits,sachdeva2020tuning,Murugan2021-eg}. 
While inverse materials design has sought to optimize specific properties \cite{rechtsman2006designed,Zeravcic2014-yb,Coli2022-qy, Zunger2018-wp,Murugan2019-ls,Lipson2000-ar,Bianchi2012-an,Engel2015-fs, Zhang2005-wt, Bianchi2012-an}, less attention has been given to identifying general design strategies for creating materials with meta-properties. 

Here, we show how a biologically-inspired design method can target one such meta-property, adaptability. By adaptability, we mean the ability to switch between mutually-incompatible functions with minimal changes in design parameters. For example, an adaptable elastic network could switch from a negative Poisson ratio to a positive one with minimal network changes, even though a single network alone can only have one Poisson ratio. In this example, the mutually-incompatible functions are the different Poisson ratios, and the design parameters are the stiffnesses of the network bonds. A truly adaptable material will be as good as a non-adaptable material at any given function, but will require fewer modifications to produce a distinct, incompatible function. 

At first glance, the existence of a truly adaptable material seems highly improbable.
However, if the design space of the material is high-dimensional, then we should generically expect that there are many distinct choices of design parameters with equivalent performance for a given function\cite{Murugan2019-ls,kashtan2005spontaneous,Greenbury2022-sd, Van_Nimwegen2006-yz, Ebner2001-rs,rocks2020revealing,rocks2021hidden}. Our goal is to identify the much more rare subsets of design solutions which both perform the given function and are adaptable.

\begin{figure*}
\begin{centering}
\includegraphics{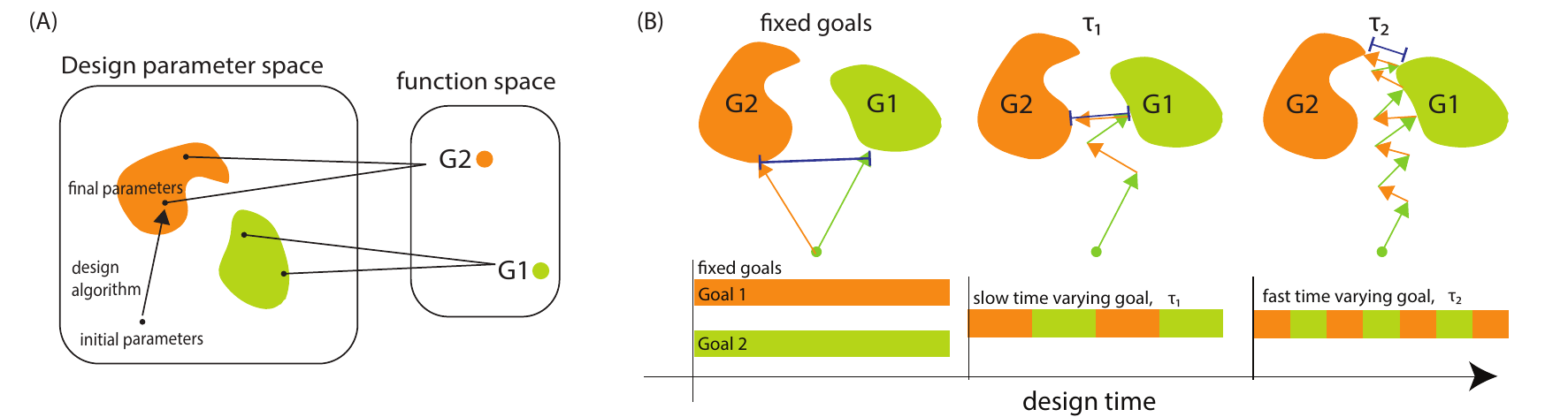}
\par\end{centering}
\centering{}\caption{\textbf{Evolving design parameters towards different target functionalities over time can select for highly adaptable materials.} (a) Materials can be characterized by their design parameters (e.g. for an elastic network, the rest length and stiffness of each bond are design parameters). The design parameter-to-function map is often highly degenerate; many choices of design parameters (e.g., orange region) can achieve the same equivalent function (e.g., $G2$). Some parts of orange region might be close to design parameters (green) that achieve an alternative function $G2$; such regions can often be vanishingly small but correspond to materials that can quickly adapt from exhibiting function $G1$ to $G2$ with minimal changes. We have drawn regions for each goal as simply connected regions, but the topology may be more complex. (b) Optimization for a single, fixed goal either $G1$ or $G2$ will typically result in non-adaptable materials that are good at function $G1$ or $G2$ but not adaptable. Switching between the two goals on a fixed timescale $\tau$, even if the parameters are not yet well-adapted to the current target goal, by construction selects for parameters that are closer together in parameter space. Faster switching selects for closer parameter sets that might be rarer or might not exist. }
\label{fig:algorithm_schematic}
\end{figure*}

In our approach, we take existing optimization algorithms for a target function and repeatedly switch the target before optimization is completed for any one function.
The partially adapted design parameters for one function are used as initial conditions for optimizing the second function. 
This intuitively requires the solutions identified in successful periods of training to drift closer to each other in design space with each switch (Fig. \ref{fig:algorithm_schematic}).

The underlying logic of this approach is that the sets of design parameters which survive the oscillating selection process are required to be similar by construction.
This implies that there are shared design characteristics between the solutions, even though the functions they perform are incompatible.

We note that our approach does not directly optimize a metric of adaptability. 
Instead, material adaptability arises because of the sequence of selection pressures the material is subject to during design optimization. Our method functions as a wrapper to existing optimization algorithms. 
It is therefore compatible with a wide range of pre-existing materials design procedures, ranging from fully computer-based\cite{rechtsman2006designed,Coli2022-qy} to fully \textit{in situ}\cite{pashine2019directed,pashine2021local}.


Our work extends intuition developed earlier on modularity\cite{kashtan2005spontaneous,hemery2015evolution} in biological contexts to canonical synthetic materials platforms. However, in these prior works, the `environment' itself was chosen to be modular.
The resulting system reflected the modules specified by the environment: e.g., with logic circuits\cite{kashtan2005spontaneous}, the environment switched between selecting for computing
an AND operation between two sub-goals, or an OR operation between those same two sub-goals.
Consequently, the resulting logic circuits developed modules for computing these sub-goals which could be quickly recombined to achieve AND or OR with minimal changes to the circuit.

In contrast, in many problems
relevant to materials, the different goals or functions required may not have any obvious modular structure.
For example, consider two goals $G_1,G_2$ representing a material with different Poisson ratios, or a polymer folding into two distinct structures with no common sub-structures. 
In this work, we focus on such arbitrary goals that are not chosen to be modular in any obvious sense. 
We will nevertheless use the alternating selection paradigm of prior biological works; we find that  such design protocols can \emph{reveal} adaptable organization of materials that can be rationalized in retrospect, even for goals not organized in any obvious modular manner. 
We demonstrate the utility of this method in the context of two distinct simulated systems - (a) soft materials with locally tunable elastic modulii, and (b) self-assembling heteropolymers with tunable interaction affinities between monomers.
In both systems, rapidly oscillating training goals allows us to find design parameters which can switch between mutually exclusive functions with minimal parameter changes.
In the self-assembly case, we gain physical insight into the origin of adaptability, as selecting for adaptability localizes parameter changes to interaction units which control kinetic barriers in the folding landscape.
Similarly, in the elastic networks, we find that adaptability arises from a coherent displacement unit which is easily shifted to perform opposing allosteric motions. Thus, our work suggests a broad strategy to discover physical mechanisms that enable adaptability in materials with arbitrary underlying physics.

\section*{Results}

\subsection*{Mechanical metamaterials}

\begin{figure*}
\begin{centering}
\includegraphics{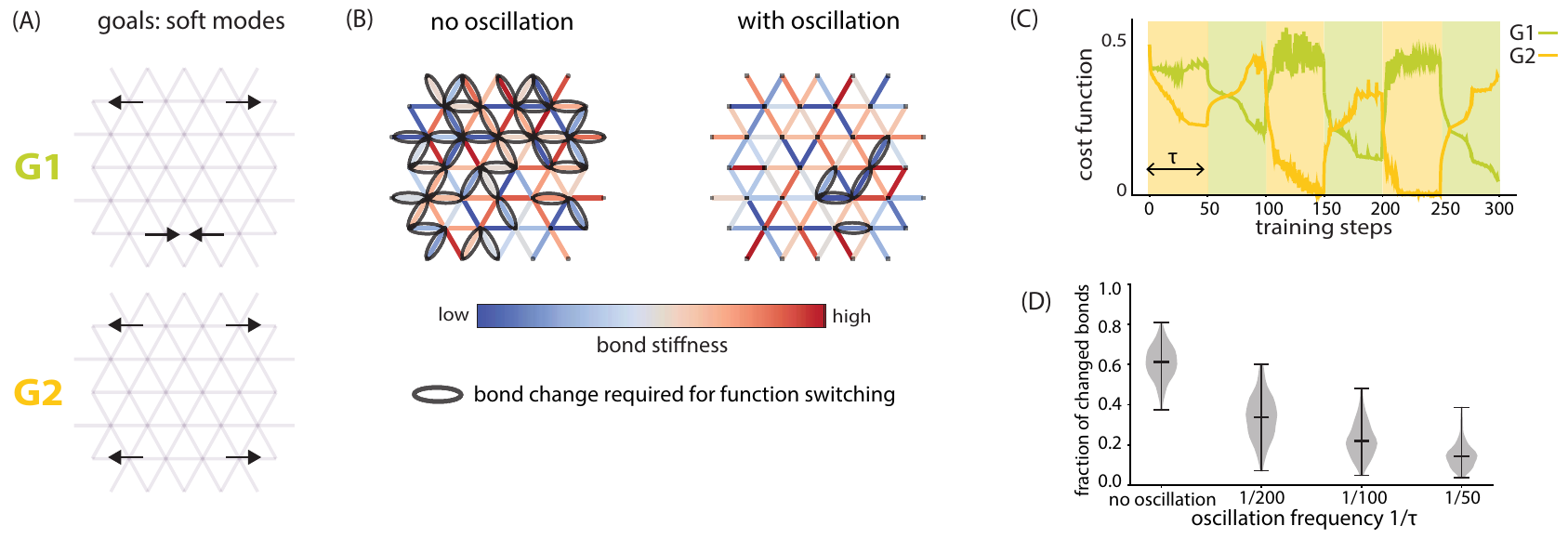}
\par\end{centering}
\centering{}
\caption{\textbf{Oscillatory training protocols generate adaptable solutions for allostery in elastic networks}
(A) We perform training on elastic networks with tunable bond stiffnesses. 
Goals $G1$ and $G2$ seek extensile and contractile strains respectively at a pair of target nodes in response to strain at a pair of input nodes. Target nodes, applied strain, and input nodes are the same for $G1$ and $G2$.
(B) Each successful run of training generates a pair of elastic networks; one performs $G1$, the other performs $G2$.
We show the network which performs $G1$, bonds colored by relative stiffness. Bond stiffnesses which change $> 10\%$ to switch to $G2$ circled in black.
Network examples from training without oscillations (left), and
with oscillations (right).
(C) Performance on each goal $G1, G2$ quantified by a cost function. Lower cost function indicates improved performance. Training drives cost function of on-target goal lower during each period. Background color panels indicate on-target goal. $\tau$ indicates training steps per period of goal oscillation.
(D) Faster oscillation (smaller $\tau$) during training gives networks with higher adaptability (defined as fraction of all bond stiffnesses which change beyond $10\%$ threshold when switching between $G1$ and $G2$). 
Violin plots show distribution of changed bond fraction over successfully trained network pairs, black lines indicate minimum, mean, and maximum values.}
\label{fig:softmode_training}
\end{figure*}


In the context of mechanical materials, we first focus on allosteric response, that is, the ability of an elastic network to exhibit a desired strain at a distant target site if strained at a specific source site.  Such allosteric responses have been created through multiple methods \cite{rocks2017designing,yan2017architecture,yan2018principles,tlusty2017physical,flechsig2017design,hemery2015evolution}.
Typically, there are multiple design parameters that give rise to an allosteric response, which means that the design space is degenerate with respect to that particular allosteric motion. 
Our goal is to search through this degenerate design space for the smaller set of parameters that can perform a specific, different, incompatible allosteric response with minimal adjustment.

We use a 2-d mass-spring network to model allostery in a mechanical system. Specifically, we simulate a 22-node hexagonal lattice with fixed boundary conditions. While the geometry and rest lengths of all springs are fixed, the collection of 83 spring constants $K = \{ k_i\}_{i=1}^{83}$ are design parameters that can be tuned to get specific allosteric responses.  The two motions we train for, $G1$ and $G2$, correspond to two opposite responses for the same input at the same source site (Fig. \ref{fig:softmode_training}A). 
We first reproduce such single-function design by starting from many random initial conditions and running a gradient descent procedure on a fitness function related to the dynamical matrix. 

Our fitness function rewards the softest mode transmitting strain from source to target while also rewarding a large energy gap between such a soft mode and the next mode; see Supplementary Information for further detail. Other works have created such allosteric materials using a range of different fitness functions and optimization procedures\cite{rocks2017designing,yan2017architecture}. 
Our procedure, while different in detail from previous work, nevertheless consistently produces networks with the desired allosteric response.

Crucially, we find that there are numerous choices of design parameters $K$ - here, bond stiffnesses - that can separately perform each of the goals $G1$ and $G2$. One of these networks for $G1$ is shown in Fig. \ref{fig:softmode_training}B(left); the others are very different in the choice of $K$ but are equivalent in terms of performance at $G1$. 



\begin{figure*}
\begin{centering}
\includegraphics[width=0.98\textwidth]{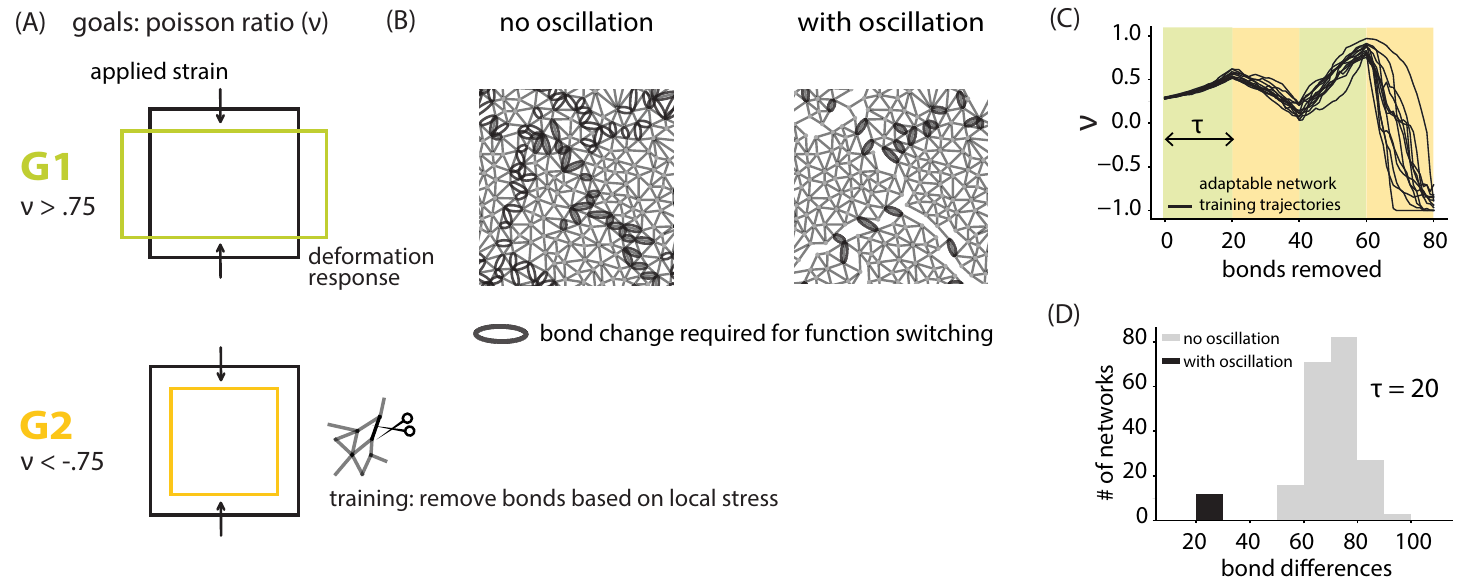}
\par\end{centering}
\centering{}
\caption{\textbf{Oscillatory training protocols generate networks with adaptable Poisson ratios.} (A) 2D elastic networks with disordered geometry are trained for Poisson ratios $\nu> 0.75$ ($G1$) or $\nu < - 0.75$ ($G2$).
During training, bonds are irreversibly removed based on local stress in response to applied strain.
(B) Each successful run of training generates a pair of elastic networks; one network performs $G1$, the other performs $G2$.
We show the network which performs $G1$. Bond changes required for switching to $G2$ are circled in black.
Network examples from training without oscillations (left), and with oscillations (right).
(C) An ensemble of $200$ networks undergoing oscillatory training with bond removal results in successful switching between $G1$ and $G2$ in 12 networks. Background color panels indicate on-target goal. $\tau$ indicates bonds removed per period of goal oscillation.
(D) Successful networks trained with oscillation have a bond difference of $20$ bonds. Networks trained without oscillation showed bond differences ranging from $50-100$.
}
\label{fig:poissonratio_training}
\end{figure*}

To leverage this degeneracy in design parameters for adaptability, we studied a family of algorithms in which the target goal is switched periodically between $G1$ and $G2$ during design optimization at different timescales $\tau$,
where $\tau$ is the number of optimization steps per period. In this way, design parameters partially optimized for one goal, say $G1$, are used as initial conditions for the next period of design that targets $G2$. 
Following our intuition in Fig. \ref{fig:algorithm_schematic}, any two sets of design parameters produced consecutively in this way are likely to be similar.

The results of switching at different frequencies are shown in Fig. \ref{fig:softmode_training}B-D. We find that such oscillatory
design has two distinct phases. Initially, the design parameters are not good at either goal $G1, G2$ at the time of training goal switches (Fig. \ref{fig:softmode_training}C). 
After this phase, we find one of two outcomes:  (a) success, i.e., convergence to a limit cycle between a pair of design parameters (bond stiffnesses) $K_1$ and $K_2$ that are good at $G1$ and $G2$ respectively (Fig. \ref{fig:softmode_training}C) or (b) failure, i.e., convergence to design parameters that are good at neither property.


In successful cases, we can measure adaptability of the pair $K_1,K_2$ as the number of bonds that need to change their elastic constant $k_{ij}$ by more than $10\%$.

To understand the robustness of our procedure, we repeated the above simulations for $500$ random initial assignments of bond stiffness for each of $4$ different oscillation timescales $\tau$ (Fig. \ref{fig:softmode_training}D).
We see a substantial and systematic increase in adaptability with frequency of switching, when restricting to successful runs. Intuitively, for more rapid switching times, when the process does converge on a limit cycle, the pair $K_1,K_2$ are closer, as they must be since there is less design time to get from one to another. 

For example, using an oscillation timescale $\tau = 50$, we discover networks that can switch function by changing as few as  $5$ bonds (Fig. \ref{fig:softmode_training}B(right)). 
In contrast, networks obtained by optimizing for $G1$ or $G2$ alone typically differ significantly in $40$ bonds (Fig. \ref{fig:softmode_training}B(right),D). 

To better understand how adaptability arises under oscillatory training, we optimized several larger networks.
While in small networks it was more difficult to identify physical principles in the optimized networks, in larger networks the mechanical signatures of allostery were visually clear.
In one example network, we found that oscillatory training produced a section of the network which moved coherently (Supplemental Fig. 1, bottom).
This section is diverted with just a small number of bond stiffness changes, shifting from an in-phase to an out-of-phase motion.
Surprisingly, training for only one motion at a time also produced networks with coherent motions  (Supplemental Fig. 1, top).
However, the coherent motions selected for in the non-oscillatory training differed between the $G1$ and $G2$ goals.
This contrast between oscillatory and non-oscillatory pairs was observed in several other large networks (Supplemental Fig. 2).

\subsubsection*{Local learning rules}

\begin{figure*}
\begin{centering}
\includegraphics[width=0.98\textwidth]{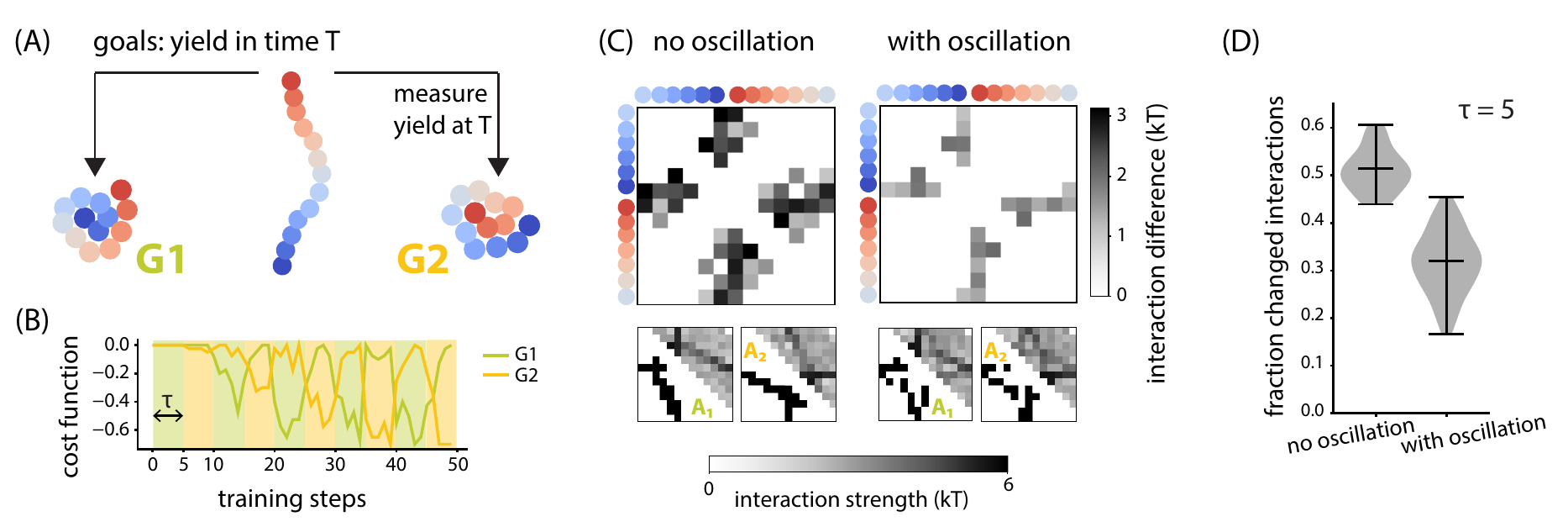}
\par\end{centering}
\centering{}\caption{\textbf{Oscillatory training protocols generate adaptable solutions for heteropolymer folding.} (A) We train the monomer interaction matrix for a heteropolymer of length $13$ in order to target different folded structures - a clockwise spiral $G1$ and a counter-clockwise spiral $G2$ - in finite time, starting from an unfolded state. Monomers are colored according to position. 
(B) Performance on each goal $G1, G2$ is quantified by a cost function. Lower cost function indicates improved performance. Training drives cost function of on-target goal lower during each period. Background color panels indicate on-target goal. $\tau$ indicates training steps per period of goal oscillation.
(C) Trained interaction matrices that target a spiral $G1$ and an anti-spiral $G2$: 
(bottom) matrices $A_1$, $A_2$; upper triangle is the matrix, and the lower triangle shows upper quartile interaction values. (top) matrix difference $|A_1 - A_2|$. Both top and bottom panels are averaged over no-oscillation (n=40) and with-oscillation (n=62) training runs, thresholded at 1 kT. Polymer ends are positioned at the center of interaction matrices.
(D) Fraction of interactions which change by $>2$ kT to switch between $G1, G2$. Violin plots show distribution over no-oscillation training pairs (n=40) and with-oscillation pairs (n=62). Lines indicate minimum, mean, and maximum values.
}
\label{fig:polymer_training}
\end{figure*}

In Fig. \ref{fig:softmode_training}, we showed that adaptable allosteric response can be created in elastic networks by alternating training for two incompatible motions. 
This training relied on the optimization of a global cost function under gradient descent. 
However, recent examples of elastic network training aim to change bulk elastic moduli with algorithms which use local information as input, and modify the network in an irreversible fashion\cite{goodrich2015principle,reid2018auxetic}. 
When training for adaptability in mechanical allostery, we implicitly assumed that the dynamics of training would allow for returns to previously visited regions of design parameter space.
It is not clear that we can train for adaptability without the ability to move unrestricted through design parameter space.
Here, show that we can extend our oscillatory training framework to the task of developing bulk elastic response even with irreversible local update rules.

Our two target goals $G1$ and $G2$ now correspond to having Poisson ratios of $>.75$ and $< -.75$ respectively (Fig. \ref{fig:poissonratio_training}A). 
We will use the same notation as in the previous section - $G1$ and $G2$ - to refer to these goals.
The Poisson ratio $\nu$ of a material describes its bulk deformation response to a uniaxial strain. 
If the applied strain is compressive, a negative $\nu$ indicates that a network will contract along the axis orthogonal to the strain, while a positive $\nu$ indicates that a network will expand along the orthogonal axis.
Note that for isotropic materials, $\nu$ is constrained to be within $[-1, 1]$, but here we consider materials which may become anisotropic as they are trained; see Supplementary Information for further detail.

Our algorithm for training elastic networks to perform $G1$ and $G2$ proceeds by the irreversible removal of bonds based on local information.
Our design parameters are the presence or absence of a bond in the lattice, but we do not allow bond additions.
We initialize our training with disordered 2D mass-spring networks of approximately 200 nodes.
We simulate these lattices under periodic boundary conditions.
During training, we enforce a deformation on the network, measure the strain in each bond, and then remove the bond which experiences the most strain.
When we train for positive $\nu$, we compress the network along the y-axis and stretch it along the x-axis.
Analogously, we train for negative $\nu$ by compressing the network along the y-axis while also compressing it along the x-axis.
During oscillatory training, we alternate which of these deformations is applied.
See Supplementary Information for further detail.

Despite the differences between the goals and algorithms considered here compared to those utilized in the design of adaptable allostery, we find that a similar picture of adaptable mechanical design under oscillatory training emerges (Fig. \ref{fig:poissonratio_training}B-D).
Even when trained from the same initial disordered network, training for $G1$ and $G2$ produce pairs of networks with many differences in their bond removals (Fig. \ref{fig:poissonratio_training}B(left)).
In contrast, when we oscillate which deformation is applied once every 20 bond removals, we find that the difference between such network pairs is substantially lower (Fig. \ref{fig:poissonratio_training}B(right)).
This alternation comes at a cost; in an ensemble of networks undergoing oscillatory training, approximately $50\%$ experience mechanical failure before the training ends, due to the irreversible nature of bond removals.
Of those that survive the training, $12\%$ are able to rapidly switch between $G1$ and $G2$, with an overall yield of $6\%$ (Fig. \ref{fig:poissonratio_training}C).
When comparing an ensemble of networks trained with oscillation to an ensemble of networks trained without, we find quantitative evidence that adaptability increases with oscillatory training, at the cost of lower yield (Fig. \ref{fig:poissonratio_training}D).

\subsection*{Heterpolymer folding}



Having demonstrated our method's success in designing adaptable mechanical networks, we turned to another paradigmatic class of tunable synthetic materials.

Programmable self-assembly of single target structures has been  explored in many systems, ranging from colloids to proteins and DNA. Across these diverse systems, a similar set of design parameters are tuned to target assembly of a desired structure. Typically, these parameters include a matrix of binding affinities between building blocks, in addition to global parameters like temperature and concentrations. We refer to the matrix of binding affinities as the affinity matrix.

In most approaches to self-assembly \cite{go1983theoretical,pande2000heteropolymer}, 
the affinity matrix closely resembles the contact matrix of the building blocks in the desired structure $\Gamma$. That is, particles in contact in $\Gamma$ should typically have stronger binding affinities compared to particles not in contact in $\Gamma$, thereby energetically stabilizing the structure relative to other configurations of the same particles.

As a result, design parameters optimal for assembling a structure $\Gamma_a$ would not be good at assembling an unrelated structure $\Gamma_b$ with high yield.
This makes adaptability in self-assembly seem difficult from the outset.
The stochastic nature of self-assembly provides an additional complication compared to elastic networks.

To test whether we can design a self-assembling system to be adaptable, we built a simulation of 2-d heteropolymer folding using the HOOMD-blue software\cite{anderson2020hoomd}. Specifically, we consider a polymer of 13 monomers, each of which is bonded to the next with harmonic springs. A harmonic bending energy is present to stabilize the fully-extended polymer state with a persistence length of 5.5. Each monomer interacts with all non-neighbor monomers through an attractive Morse potential with a tunable affinity. The affinities have a maximal value of 6 kT and a minimal value of 0 kT. As such, the design space has dimension 66, equal to the number of lower-triangular entries in a $13 \times 13$ affinity matrix without including the first off-diagonal; see Supplementary Information for further simulation details.

Our task is to select affinity matrices which can readily switch between two distinct, mutually exclusive design goals (Fig. \ref{fig:polymer_training}A). Our first goal, $G1$, is to produce polymers that fold from a fully-extended initial condition into a spiral with a winding number exceeding that of 1.0 around the monomer at the head of the polymer. As self-assembly is stochastic, we further specify that this must occur with greater than 70\% probability within $500$ units of simulation time. $G2$ is defined analogously, with the difference being that the target structure is now an anti-spiral, which has a winding number exceeding that of 1.0 measured relative to the monomer at the tail of the polymer.
Note that use the same notation as in previous sections - $G1$ and $G2$ - to refer to these goals.



We optimize the yield of a given target structure over the 66 design parameters 
using the Covariance Matrix Adaptation-Evolutionary Strategy (CMA-ES) \cite{hansen1996adapting,hansen2006cma} that simulates an evolving population of design parameters.
The loss function for our implementation of CMA-ES is the negative of the yield, with a floor set by the minimum $70\%$ yield required for successfully achieving either $G1$ or $G2$; see Supplementary Information for further parameter choice details.

We perform optimization with two training protocols: 1. ``no-oscillation'' training, where $G1$ and $G2$ are optimized individually, and 2. ``with-oscillation'' training, where we switch between $G1$ and $G2$ with a period of $5$ training steps. When we successfully perform optimization in with-oscillation training, we see that the maximal yield of each goal increases in an alternating fashion with each passing training period (Fig. \ref{fig:polymer_training}B).
This suggests that oscillating training is converging to affinity matrix solutions which can easily switch between $G1$ and $G2$.
Through this procedure, we collect a set of affinity matrix pairs $A_1, A_2$. 
Additionally, we verify that the adaptability of our pairs $A_1, A_2$ do not come at the expense of significant performance degradation on the individual goals they are trained for (Supplemental Fig. 3).
We collect an analogous set of pairs for no-oscillation training simply by running converged optimizations from the same initial conditions.

To characterize the distribution of successful affinity matrix pairs, we computed the average difference between a matrix that achieved $G1$ and its corresponding $G2$ partner (Fig. \ref{fig:polymer_training}C, top row), focusing only on entries that changed substantially (i.e., by more than $1$ kT). 
The resulting average difference matrix from the no-oscillation training shows many more changed entries than the matrices from with-oscillation training.
This visually suggests that oscillatory training identifies more adaptable regions of design space, where the affinity matrices which achieve $G1$ are closer to those which achieve $G2$. The average affinity matrix for each goal supports the same conclusion  (Fig. \ref{fig:polymer_training}C, bottom row). 
Note that we have shifted the monomer numbering when plotting affinity matrices, for ease of visual comparison.

To quantitatively confirm these visual conclusions, we compute the fraction of matrix entries that change by more than $2$ kT between each $A_1, A_2$ affinity matrix pair (Fig. \ref{fig:polymer_training}D). As expected, the distribution of this metric across all such pairs is substantially higher for no-oscillation training than for with-oscillation training.

\subsubsection*{Physical interpretation}

\begin{figure}
\begin{centering}
\includegraphics[width=0.4\textwidth]{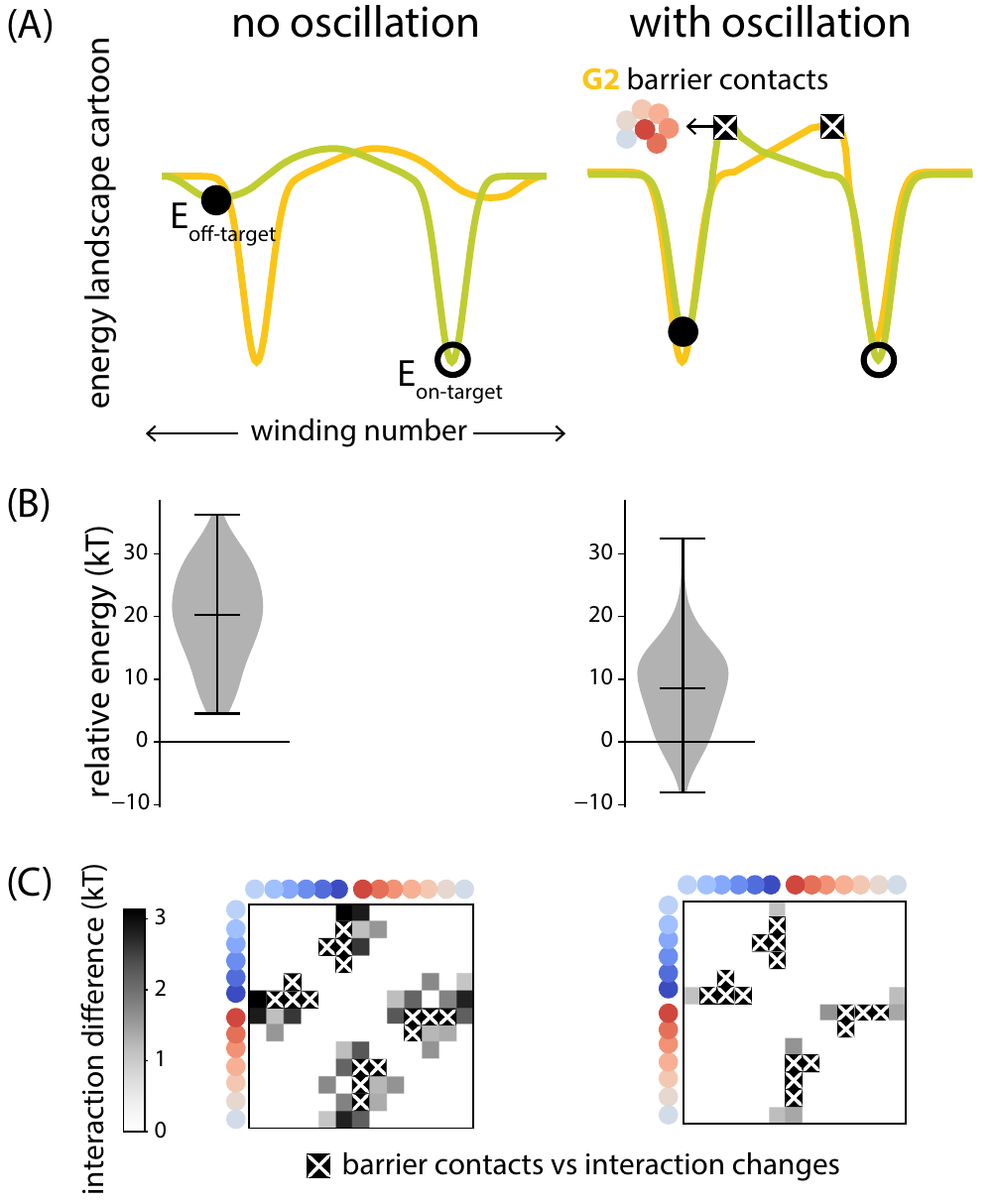}
\par\end{centering}
\centering{}\caption{\textbf{Adaptability in polymer folding relies on localizing interaction changes to nucleation barriers.} A. When trained without oscillations, energy landscapes (cartoons) targeting $G1$ (green) and $G2$ (orange) have deep energy minima at their on-target $G1$ and $G2$ (black open circle) but high energy at off-target $G2$ and $G1$ (black solid circle) respectively. With oscillatory training, discrimination is now kinetic; both on- and off-target energies remain low, but the landscape develop kinetic barriers to folding (white cross black squares). 
Kinetic barriers enable adaptability since folding can changed by changing a small number of contacts in the nucleation seed. 
B. Off-target energy distributions for no-oscillation (left) and with-oscillation (right) training (data from 40 and 62 simulation runs respectively); energy is relative to the mean of the no-oscillation on-target distribution. The off-target distribution is lower in energy than in with-oscillation training. Violin plot lines indicate minima, means, and maxima.
C. As in Fig. \ref{fig:polymer_training}C, interaction matrix differences between $G1$ and $G2$, for no-oscillation (left) and with-oscillation (right) training.  Barrier contacts are overlaid as white cross black squares; oscillatory training localizes interaction matrix changes to the barrier contacts. Monomers are colored according to position.}
\label{fig:nucleation_interpretation}
\end{figure}


The adaptability of self-assembly found here is surprising at first glance. The average adaptable pair of affinity matrices $A_1,A_2$ resemble each other for the majority of elements (Fig. \ref{fig:polymer_training}C, right), yet fold into incompatible configurations with high yield.  

To understand the physical design principles underlying such adaptability, we estimated aspects of the folding energy landscape for affinity matrices $A$ found through oscillatory and non-oscillatory training; we computed the energies of folded configurations with different winding number (see Supplementary Information).
We find that on-target structures are similarly stabilized by both oscillatory and non-oscillatory training as suggested by the cartoon in Fig. \ref{fig:nucleation_interpretation}A (open circles). 

However, the two training protocols differ in how they treat off-target structures. With non-oscillatory training, the off-target structures are relatively high in energy (solid circle in Fig. \ref{fig:nucleation_interpretation}A,B) since training is only ever shown the on-target structure. Consequently, the affinity matrix requires extensive changes to assemble the previously-off-target structure (Fig. \ref{fig:nucleation_interpretation}C).

In contrast, with oscillatory training, off-target structures remain low energy states even with the affinity matrix that target a radically distinct on-target structure. Relative to the no-oscillation off-target distribution, the with-oscillation off-target distribution is lower by $\sim 10$ kT (Fig. \ref{fig:nucleation_interpretation}B). 

Despite such energetic stabilization of off-target structures, with-oscillation training results in on-target folding by exploiting kinetics. Folding is controlled by a nucleation barrier that is higher for the off-target structure than for the on-target structure (cartoon Fig.\ref{fig:nucleation_interpretation}A; right). 

Using the estimated energy landscapes, we can identify ``barrier contacts'' that need to form in a partially folded nucleation seed before subsequent downhill folding to completion. 
Oscillatory training localizes the few changed affinities to those involved in forming barrier contacts (black in Fig. \ref{fig:nucleation_interpretation}C).



Thus, our time-varying algorithm points at a physical principle for adaptive self-assembly of independent validity. Kinetic yield is controlled by partially folded early intermediate structures that correspond to nucleation barriers. These barriers can be lowered in energy or conversely destabilized by relatively few changes to the affinity matrix, resulting in the spiral or anti-spiral with high selectivity. Similar principles might apply more broadly to proteins and ribozymes where partly folded configurations, \textit{en route} to fully-folded configurations can be destabilized; indeed, such mechanisms might operate in experimentally characterized adaptable proteins and ribozymes where a single mutation can switch the polymer between distinct structures and thus function.

\section*{Discussion}

Biological materials differ from man-made materials in not just their physics and composition but also in the history of their development. In fact, the way biological materials are arrived at, through a process of incremental evolution in a sequence of historic environments, is critical in understanding why they function differently from man-made materials. 
While man-made materials can sometimes rival or even exceed specific functionalities of natural systems, 
these synthetic systems are lagging precisely in meta-properties like adaptability, robustness and ability to acquire new functions on the fly.  

Here, we have shown how one such property, adaptability, can arise without any direct targeting or optimization.
Instead, we find adaptable materials by applying a time-varying sequence of selection pressures during design optimization. 

This adaptability comes through the spontaneous formation of identifiable physical units, despite the fact that our goals and systems were not explicitly modular in form.
In contrast, prior work\cite{kashtan2005spontaneous,hemery2015evolution} used explicit modularly varying goals to show a potential origin of modularity and adaptability in biological systems. Thus our work can be seen as building on those ideas to discover physical principles specific to the physics of systems studied (e.g., nucleation for polymer folding) that allow for adaptability.


Our proposed method has potentially wide applicability, in that it functions as a wrapper around pre-existing design programs, and hence can be applied without in-depth knowledge of a system.


However, our method can also help reveal new system-specific physical insights 
that can then be exploited without further need for our method. 
For example, in many of the current platforms for self-assembly, the yield is frequently governed by kinetics rather than equilibrium free energies\cite{jacobs2015rational,hensley2022self,bitran2020cotranslational}.
Our simulations of heteropolymer self-assembly revealed a broadly relevant design principle for such systems -- nucleation barriers in energy landscapes can be leveraged to create adaptability in self-assembly.
Similarly, in the context of elastic networks, we identified that coherent motions which link two allosteric sites can be easily diverted in order to achieve incompatible goals.
These insight can now be used to guide design without need for blind numeric optimization, both in synthetic systems like colloids, DNA, and foams, 
but also in natural systems like proteins\cite{bitran2020cotranslational}.


One key condition for our method is already clear from the pilot examples in this study. 
In order for our adaptability framework to succeed, the goals under consideration must have ``neutral variation''\cite{Gruner1996-ll, Gruner1996-dj,Schultes2000-mg}; there must exist changes in design parameters that have no cost in terms of the current target functionality but that help adapt to new functionality.
The systems studied here have such degeneracy; for example, many elastic networks with different bond stiffnesses showed the same desired allosteric response. In fact, degeneracy is generically expected whenever systems are disordered, with the number of design parameters often being extensive in the size of the system.  
In biological examples too, such genotype-to-phenotype maps are often redundant when the space of genotypes (or design parameters) is larger dimensional than the space of phenotypes (exhibted properties). We do not expect as much success in systems such as self-assembled crystals with only 1 or 2 species or regular lattices of elastic elements.







\textbf{Acknowledgments.}
The authors thank Varda Hagh, Heinrich Jaeger, Ilya Nemenman, Rama Ranganathan, Riccardo Ravasio, and Kabir Husain for discussions. This work was primarily supported by the University of Chicago Materials Research Science and Engineering Center, which is funded by the National Science Foundation under award number DMR-2011854. Arvind Murugan acknowledges support from the Simons Foundation. M.L.G. acknowledges support from NSF Grant DMR- 2215605. Ayanna Matthews acknowledges support the University of Chicago Biophysics Training Grant.

\bibliographystyle{aiaa}
\bibliography{osc.bib,Paperpile.bib}

\end{document}